\documentstyle[11pt]{article}

\newcommand{\eps}{\varepsilon}
\newcommand{\eq}[2]{\begin{equation}\label{#1} #2 \end{equation}}
\newcommand{\eqs}[2]{\begin{equation}\label{#1}\begin{array}{rcl}
                     #2 \end{array}\end{equation}}
\newcommand{\equthree}[2]{\begin{equation}\label{#1}\begin{array}{rcrcl}
                     #2 \end{array}\end{equation}}
\newcommand{\eqar}[2]{\begin{eqnarray}\label{#1} #2 \end{eqnarray}}
\newcommand{\Jop}{\hat{J}}
\newcommand{\vop}{\hat{v}}
\newcommand{\xop}{\hat{x}}
\newcommand{\dvec}{\vec{d}}
\newcommand{\kvec}{\vec{k}}
\newcommand{\Rvec}{\vec{R}}
\newcommand{\kp}{\vec{k}_{\|}}

\newcommand{\GL}{G^<}
\newcommand{\GG}{G^>}
\newcommand{\ga}{\raisebox{-.5ex}{$\stackrel{>}{\sim}$}}
\newcommand{\la}{\raisebox{-.5ex}{$\stackrel{<}{\sim}$}}
\newcommand{\tr}{\mbox{tr}}
\renewcommand{\thesection}{\Roman{section}}

\renewcommand{\theequation}{\arabic{section}.\arabic{equation}}

\begin{document}

\vspace{.1in}

\begin{center}  \begin{bf}
\begin{large}
{Multiband tight--binding approach to tunneling in semiconductor
heterostructures: Application to $\Gamma X$ transfer in GaAs}
\end{large}

\vspace{0.2in}
J. A. St\o vneng$^*$
\end{bf}

NORDITA, Blegdamsvej 17, DK--2100 Copenhagen \O, Denmark

\begin{bf}

\vspace{0.1in}
and

\vspace{0.1in}
P. Lipavsk\'{y}
\end{bf}

Institute of Physics, Academy of Sciences of Czech Republic,

Na Slovance 2, 18040 Praha 8, Czech Republic

\end{center}

\vspace{0.25in}
\noindent
We study tunneling in semiconductor heterostructures
where the constituent materials can have a direct or indirect bandgap.
In order to have a good description of the lowest conduction band, we
have used the nearest--neighbour $sp^3s^*$ tight--binding model put
forward by P. Vogl {\em et al.}. A recursive Green--function method
yields transmission coefficients from which an expression for the
current density may be written down.
The method is applied to GaAs/AlAs heterostructures. Electrons
may traverse the AlAs barriers via different tunneling states
$\psi_{\Gamma}$ and $\psi_X$ ($\Gamma X$ mixing). With an applied
bias $V\ga0.5$ V electrons may enter the GaAs collector contact in
both the $\Gamma$ and the $X$ valley ($\Gamma X$ transfer). We have
studied a number of GaAs/AlAs structures. For very
narrow barriers there is little $\Gamma X$ transfer, but AlAs barriers
wider than about 25 \AA\ act as ``$\Gamma X$ filters'', i.e., most
transmitted electrons have been transfered to the $X$ valley.

\newpage
\setcounter{equation}{0}
\section[Intro]{$\!\!\!\!\!\!\!.\,\,$ INTRODUCTION} \label{sec.intro}

\hspace{\parindent}
Tunneling in semiconductor heterostructures has attracted considerable
interest over the last decade. A very important motivation factor has been
the progress in advanced crystal--growth techniques like molecular--beam
epitaxy (MBE). The ability to grow nearly perfect layered structures has
enabled experimental verification of predictions based on relatively
simple theoretical models. There is also a great interest in making
electronic devices based on such structures.

Most treatments of transport through heterostructures have been based on
effective--mass theory. This is a good approximation when the different
materials which form the structure all belong to the same category with
respect to type of energy bandgap, {\em direct} or {\em indirect}. An
example of a system with only direct--bandgap constituents is
GaAs/Al$_x$Ga$_{1-x}$As, provided the Al concentration is sufficiently
low, $x\la 0.4$. In a typical experiment, doping in the GaAs contacts
yields a Fermi level $E_F$ of the order of 10--50 meV above the
conduction--band minimum $E_c^\Gamma$ at the $\Gamma$ point of the
Brillouin zone $(\kvec=0)$. Since the conduction band is nearly
parabolic in the range from $E_c^\Gamma$ to $E_F$, all the incoming
electrons are well described by a single effective mass $m_\Gamma^*$(GaAs).
Furthermore the lowest tunneling barrier is determined by the
conduction--band minimum in Al$_x$Ga$_{1-x}$As,
which is also at the $\Gamma$ point.
Thus the {\em tunneling states}, through which electrons with energy $E$
can traverse the barrier region, are characterized by an imaginary
wave vector $k=i\kappa$, where $\kappa$ is determined by the
tunneling--barrier height $E_c^\Gamma$(Al$_x$Ga$_{1-x}$As)$-E$
and the effective mass $m_\Gamma^*$(Al$_x$Ga$_{1-x}$As).

For Al concentrations $x>0.4$, Al$_x$Ga$_{1-x}$As becomes an
indirect--bandgap material. The valence--band maximum
is still at the $\Gamma$ point, but
the conduction--band minimum is now $E_c^X$, close to the $X$ point,
$\kvec=\frac{2\pi}{a_L}(100)$, at the edge of the Brillouin zone. Here
$a_L$ is the lattice constant of the zincblende material. When
indirect--bandgap Al$_x$Ga$_{1-x}$As is used as the barrier
material, it is no longer
sufficient to take into account only the tunneling states which correspond to
the conduction--band minimum at $\Gamma$. Important are also the states
that correspond to the analytical continuation into the energy gap of
real--$\kvec$ states at the conduction--band minimum near the $X$ point.
``$\Gamma$ states'' and
``$X$ states'' may have comparable decay lengths since the lower barrier
height of the $X$ states, $E_c^X-E$, is compensated for by a higher
effective mass. The importance of $X$ states in tunneling through
indirect--bandgap barriers has already been appreciated in several
experiments
\nocite{hase,mendez1,couch,mendez2,skolnick,shields,lu1,pritchard,lu2,
austing}
\cite{hase}--\cite{austing}
and it has also been studied
theoretically by a number of groups
\nocite{cade,liu,xia,rossmanith,zheng,sankaran,schulz,ting}
\cite{cade}--\cite{ting}.

In the present paper we want to focus on the $X$ states in {\em both}
the barrier and the contact material. It is well known that electrons
in bulk GaAs may be transfered from the $\Gamma$ to the $X$ valley
if they are accelerated in a sufficiently strong electric field.
This is the Gunn effect which may be accompanied by a region
of negative differential conductivity in the current--voltage curve due
to the low mobility of electrons in the $X$ valley. The ``$\Gamma X$
transfer'' is usually stimulated by some kind of scattering mechanism, e.g.,
elastic--impurity scattering. Our aim is to show how a heterostructure with
one or more indirect--bandgap barriers can be used to control the $\Gamma
X$ transfer in a material like GaAs.
Let us assume that the transport takes place in the (100)
direction.  Then, with a bias
$V>[E_c^X($GaAs$)-E_F]/e$ applied across the
heterostructure, incoming electrons in the left
contact can end up in two final states in the right contact, in the
$\Gamma$ valley or in the $X$ valley. The
two outcomes are characterized by transmission coefficients $T_{\Gamma}$
and $T_{X}$, respectively, which depend on
the energy of incoming electrons, applied bias, and
barrier parameters. In Sec. V below we shall explore these dependencies for
GaAs/AlAs heterostructures in order to find effective
``$\Gamma X$ filters'', i.e., conditions under which $T_{X}\gg T_{\Gamma}$.

In transport experiments the measured quantity is usually the electric
current vs applied voltage. Based on simple arguments one can write down
an expression for the current density in terms of transmission
coefficients and a factor ensuring that initial states are occupied
and final states empty. In the present case a sum over different final
states at equal energy automatically provides a decomposition of the
current density in a ``direct'' component $J_\Gamma$ and a ``transfered''
one $J_X$. This simple procedure is indeed applicable in the present
study of perfectly layered heterostructures where we ignore effects of
disorder and inelastic scattering. If such effects are to be included, one
must resort to a more general approach. In an Appendix we derive an expression
for the current density which provides the starting point for extensions
to more realistic calculations \cite{still1particle}. The general expression
involves nonequilibrium Green functions and gives the {\em total} current
density $J$, {\em not} the decomposition into
$J_\Gamma$ and $J_X$. However, in the
case where disorder and phonons can be ignored, we will show how the general
expression for $J(V)$ can be decomposed and proven to be identical to the
result which was written down directly in terms of transmission
probabilities.

The transmitted electrons will be subject to elastic and inelastic
scattering and eventually come to thermal equilibrium somewhere in
the right contact. This process may be described by scattering rates
$\tau_{\Gamma\Gamma}^{-1}$,$\tau_{XX}^{-1}$ and $\tau_{X\Gamma}^{-1}$,
where the former two account for intravalley scattering and the latter
describes relaxation of $X$ electrons to the $\Gamma$ valley. Clearly,
$\tau_{X\Gamma}$ sets the time scale for observing or taking direct
advantage of having electrons in the $X$ valley.

The rates of elastic and inelastic scattering depend mainly on impurity
concentration and temperature, respectively. At low temperatures the
dominating inelastic process in GaAs is spontaneous emission of LO phonons.
A rough estimate, based on Refs.~\cite{adachi1,fawcett,ridley}, yields an
inelastic relaxation rate on the order of $10^{13}$ s$^{-1}$ in both the
$\Gamma$ and the $X$ valley. A similar estimate yields an {\em intervalley}
scattering rate $\tau_{X\Gamma}^{-1}\sim10^{12}$ s$^{-1}$,
taking into account that
electrons in the $X$ valley may relax to the $\Gamma$ valley via emission
of LO phonons or via scattering off charged impurities (assuming impurity
concentrations of about $10^{17}$ cm$^{-3}$). This implies that, with an
average drift velocity of about $10^5$ m/s, a transfered electron travels
typically a distance of 1000 \AA\ in the $X$ valley.

One way of detecting $X$ electrons, then, could be by means of a
magnetic field $B$ applied parallel to the crystal--growth direction, and
taking advantage of the difference in Landau--level splitting
$(\Delta E_{LL}=\hbar\omega_c=\hbar e B/m^*)$ in the $\Gamma$ and the $X$
valley due to the difference in effective mass $(m_X^*\gg m_{\Gamma}^*)$.
However, in order to test predictions for the $\Gamma X$ transfer,
it may even, for carefully designed heterostructures, be sufficient to
measure the {\em total} current vs applied voltage. If the
nonlinear structure in the measured total
current is well described by the theory, it seems reasonable to assume that
the calculated decomposition into $J_{\Gamma}$ and $J_X$ also agrees with
reality.
Finally, the $\Gamma X$ transfer may be checked by investigating the noise
spectrum of the current. Quite often the noise spectrum of a given
physical quantity may reveal sharper features and more information than
a measurement of the physical quantity itself \cite{hanke}.


The effects we want to study involve states far from the local minima of
the conduction band and also states in more than one valley. In order to
describe the band structure correctly, at least qualitatively, we apply
the semiempirical $sp^3s^*$ tight--binding (TB) Hamiltonian put forward by
Vogl {\em et al.} \cite{vogl}. A feature of this model, of particular
interest here, is the correct description of the transition in
Al$_x$Ga$_{1-x}$As from a direct to an indirect bandgap. In addition
it is a three--dimensional model which makes it suitable for
evaluation of current--voltage characteristics. The $sp^3s^*$ model
was also used by Cade {\em et al.} \cite{cade} and by Yamaguchi
\cite{yamaguchi}.

We have organized the paper as follows. In Sec. II we describe briefly the
$sp^3s^*$ TB Hamiltonian. In Sec. III we derive expressions for the
scattering coefficients. An expression for the current density is presented
in Sec. IV, and in Sec. V we perform a systematic study of
the $\Gamma X$ transfer in GaAs/AlAs heterostructures.
We conclude in Sec. VI and discuss briefly possible extensions and
other applications of the present model.

\setcounter{equation}{0}
\section[sp3ss]{$\!\!\!\!\!\!\!.\,\,$THE $sp^3s^*$ HAMILTONIAN}
\label{sec.sp3ss}

\hspace{\parindent}
The semiconductors that we want to describe have the crystal
structure of zincblende. The {\em valence bands} of these materials
are usually well described by an eight--band $sp^3$ nearest--neighbour
semi--empirical TB Hamiltonian \cite{harrison,vogl}, the
eight bands arising from the inclusion of four atomic orbitals on
each of the two types of atoms, cation and anion. However, for
electronic transport and tunneling we need a model which describes the
{\em lowest conduction bands} adequately, and for that purpose the
nearest--neighbour $sp^3$ model generally fails. In particular, as shown
by Chadi and Cohen \cite{chadi}, it cannot produce an indirect
bandgap in materials like AlAs, and this is the single most important
feature required for our purposes.

Vogl {\em et al.} \cite{vogl} have overcome this deficiency by the
{\em ad hoc} inclusion of an excited $s$ state on each atom. The main
effect of coupling these $s^*$ states to nearest--neighbour $p$ states is
to repel the $p$--like conduction--band levels near the $X$ and $L$ points
to lower energies, thus producing the desired indirect bandgap. The
resulting ten--band $sp^3s^*$ Hamiltonian has thirteen independent
TB matrix elements which are determined by fitting
band--structure data.

The independent TB matrix elements are given in a basis of
symmetrically orthogonalized atomic orbitals $|nb\Rvec)$, also called
L\"{o}wdin orbitals \cite{lowdin}. Here $\Rvec$ is the position of the
atom, $n$ denotes the type of orbital ($s$, $p$ or $s^*$) and $b$ the
type of atom ($c$ for cation and $a$ for anion). In Ref.~\cite{vogl} the
Hamiltonian for a bulk material is expressed as a $10\times10$ matrix in
a basis $|nb\kvec)$ which is obtained by a discrete Fourier transform
of the localized orbitals $|nb\Rvec)$. The systems that we want to study
are translationally invariant in the ``parallel'' plane $(y,z)$. However,
the heterostructure breaks the invariance in the crystal--growth direction
which is usually chosen to be along the (100) axis. Thus it is convenient
to represent the Hamiltonian in a basis $|nbj\kp)$ with parallel momentum
$\kp=(k_y,k_z)$ and ``perpendicular'' position $x_j=ja_L/4$ as
parameters. The distance between nearest--neighbour planes is one forth
the lattice constant $a_L$. A simple inverse Fourier transform (with the
same normalization convention as in Ref.~\cite{vogl}),
\eq{2.1}
{|nbj\kp)=L_{BZ}^{-1/2} \int dk_x e^{-ik_xja_L/4} |nb\kvec),}
yields the desired basis. Here $L_{BZ}=
8\pi/a_L$ is the length of the one--dimensional (1D) Brillouin zone
over which the $k_x$ integral is taken.

In the new basis the Hamiltonian takes the form
\eq{2.2}
{H=\left[ \begin{array}{cccccccc}
 \ddots&\ddots&\ddots& & & & & \\
   &V_{ca}&E_c&U_{ca}& & & & \\
   & &U_{ac}&E_a&V_{ac}& & & \\
   & & &V_{ca}&E_c&U_{ca}& & \\
   & & & &U_{ac}&E_a&V_{ac}& \\
   & & & & &\ddots&\ddots&\ddots
\end{array} \right],}
where each element is a $5\times5$ matrix. The matrices $E_c$ and $E_a$ are
diagonal and represent the orbital energies on cations and anions,
respectively,
whereas the ``hopping'' terms $U_{ca},U_{ac},V_{ca}$ and $V_{ac}$ involve
the various transfer--matrix elements between orbitals on neighbouring
sites. Obviously $U_{ac}=U_{ca}^{\dag}$ and
$V_{ac}=V_{ca}^{\dag}$ so that the
Hamiltonian is hermitian. In Appendix A the elements of $H$ are given in
detail, and we have also included the numerical values of the matrix
elements for GaAs and AlAs, taken from Ref.~\cite{vogl}. The Hamiltonian
in Eq.(\ref{2.2}) is formally identical to that of a two--atomic 1D chain with
interatomic separation $a_L/4$ and periodicity $a_L/2$, see Fig. 1.

Alloys will not be treated explicitly in the present paper. We only mention
that the simplest approach would be the virtual--crystal approximation (VCA).
This means that the Hamiltonian of an alloy A$_x$B$_{1-x}$C is approximated
by the weighted average of the Hamiltonians of AC and BC, e.g., $U_{ca}$[A$_x$
B$_{1-x}$C]$=xU_{ca}$[AC]$+(1-x)U_{ca}$[BC]. Scattering due to disorder
in the cation planes is neglected in the VCA. That effect can be
included by replacing VCA with the socalled coherent--potential
approximation (CPA).

Another effect which is ignored in this work, is elastic scattering due to
interface roughness. Like alloy disorder, interface roughness breaks the
translational invariance in the parallel plane. Hence the parallel momentum
is no longer conserved, and the simple treatment of the next section must
be modified. Interface--roughness scattering can also be accounted for
within the coherent--potential approximation.

\setcounter{equation}{0}
\section[Scat]{$\!\!\!\!\!\!\!.\,\,$ SCATTERING COEFFICIENTS}
\label{sec.scat}

\hspace{\parindent}
When the physical system is described with an effectively 1D
nearest--neighbour TB Hamiltonian, the scattering states can be evaluated
efficiently with a recursive Green--function technique. This method has
been applied to various problems \cite{sols}. The present treatment
will essentially be a generalization of the simple one--band 1D chain
described in detail in Ref.~\cite{stovneng}.

Since we will discuss only processes where the parallel momentum
$\kp$ is conserved, we will suppress $\kp$ in most of the notation
of this section. It is furthermore convenient to label a cation layer
and its neighbouring anion layer to the right with a common ``site''
index $j$. Assume that the heterostructure and the
undoped spacer layers, if any,
are located on the sites from $j=1$ to $j=N$. Then the homogeneous
``leads'' extend from $j=-\infty$ to $j=0$ and from $j=N+1$ to $j=\infty$.
Write the total wave function $|\Psi)$ as a sum of two pieces,
\eq{3.1}
{|\Psi)=|S_\alpha)+|\phi).}
The first term is taken to be the incoming plane--wave part with
wave number $\kvec_\alpha=(k_\alpha,\kp)$ and energy $E$, restricted to layers
up to and including the cation part of site 0:
\eq{3.2}
{|S_\alpha)=\sum_{j\leq 0,c} e^{ik_\alpha j a_L/2} |j,\alpha).}
The local part $|j,\alpha)$ is a $10\times1$ vector with coefficients
$\alpha_{nb}$ ($n=s,p_x,p_y,p_z,s^*$; $b=c,a$). These coefficients are
determined by the solution of the Schr\"{o}dinger equation for the bulk
material at energy $E$ and wave vector $\kvec_\alpha$. The problem is now
reduced to finding the ``remainder'' $|\phi)$, and since $(E-H)|\Psi)=0$,
we have
\eq{3.3}
{|\phi)=G^R(E)\left[-(E-H)|S_\alpha)\right].}
Here $G^R(E)=\lim_{\eta\rightarrow0^+}(E-H+i\eta)^{-1}$ is the
retarded Green function which also depends on $\kp$ via the Hamiltonian.
{}From the definition (\ref{3.2}) of $|S_\alpha)$, and since $H$ only couples
nearest--neighbour atomic layers, the term in square brackets in (\ref{3.3}) is
clearly zero for sites $j>0$. Furthermore the Schr\"{o}dinger equation
ensures that $(E-H)|S_\alpha)=0$ for sites $j<0$, and one is left with
nonzero terms only on the cation and anion layer at site $j=0$.
We may finally write
\eq{3.4}
{|\phi)=\frac{4i\hbar}{a_L}G^R(E)\vop|0,\alpha),}
where $\vop$ resulting from (\ref{3.3}) reads
\eq{3.5}
{\vop=\frac{a_L}{4\hbar}
\left[ \begin{array}{cc}
            0     & iU_{ca} \\
         -iU_{ac} & 0
  \end{array} \right].}
The operator $\vop$ can be interpreted as a local velocity operator. This is
consistent with the velocity operator obtained from the Heisenberg equation
of motion, $\vop=-\frac{i}{\hbar}[\xop,H]$, with the position operator
\eq{3.6}
{\xop=\sum_j \left[|j_c)\frac{ja_L}{2}(j_c| +
|j_a)(\frac{ja_L}{2}+\frac{a_L}{4})(j_a|\right].}
Here we have split the vector $|j)$ into its cation and anion parts
$|j_c)$ and $|j_a)$, respectively.

It is now straightforward to determine the scattering amplitudes. The
total wave function may be written as
\eqar{3.7}
{|\Psi)&=&\sum_{j\leq0}\left[e^{ik_\alpha ja_L/2}|j,\alpha) +
         \sum_\beta \sqrt{\frac{v_\alpha}{|v_\beta|}}\,r_{\alpha\beta}
         e^{-ik_\beta ja_L/2}|j,\beta)\right] \nonumber \\
      &+&\sum_{j\geq N+1}\sum_\beta \sqrt{\frac{v_\alpha}{v_\beta}}\,
         t_{\alpha\beta}e^{ik_\beta ja_L/2}|j,\beta) \nonumber \\
      &+&\sum_{1\leq j\leq N}\sum_\kappa \chi_{\alpha\kappa}^j |j,\kappa).}
The incoming plane wave with wave number $k_\alpha$ may be reflected or
transmitted with wave number $k_\beta$, with the restriction that the total
energy and the parallel momentum are conserved. The velocity factors are
included so that particle conservation yields a natural probabilistic
interpretation of the transmission and reflection coefficients,
\eq{3.8}
{\sum_\beta \left[|t_{\alpha\beta}|^2 + |r_{\alpha\beta}|^2\right] = 1.}
Information about the various tunneling states are contained in the
coefficients $|\chi_{\alpha\kappa}^j|^2$, the $j$ dependence of which
yields the spatial--decay rate of the state with (in general) complex
wave number $k_\kappa$. An expression for the transmission amplitude
$t_{\alpha\beta}$ is obtained by comparing the projection $\{N+1,\beta|
\Psi)$ of Eqs. (\ref{3.1}) and (\ref{3.7}):
\eq{3.9}
{t_{\alpha\beta}=\sqrt{\frac{v_\beta}{v_\alpha}}\,e^{-ik_\beta(N+1)a_L/2}
\frac{4i\hbar}{a_L}\{N+1,\beta|G^R \vop|0,\alpha).}
In a similar way the projection $\{0,\beta|\Psi)$ yields the
reflection amplitude
\eq{3.10}
{r_{\alpha\beta}=\sqrt{\frac{|v_\beta|}{v_\alpha}}\,\left[\frac{4i\hbar}{a_L}
\{0,\beta|G^R \vop|0,\alpha)-
\delta_{\alpha\beta}\{0_a,\alpha|0_a,\alpha)\right],}
where the second term comes from the fact that $|S_\alpha)$ does not include
the anion layer of site $j=0$. Eqs. (\ref{3.9}) and (\ref{3.10})
are readily seen to reduce
to the expressions for $t(k)$ and $r(k)$ given in Eq. (2.9) of
Ref.~\cite{stovneng} for the simple case of a one--band 1D model.

Note that the ``left'' states $\{j,\beta|$ in (\ref{3.9}) and
(\ref{3.10}) are not simply the hermitian conjugate $(j,\beta|$ of
$|j,\beta)$. The reason is that the local projections $|j,\beta)$ on a
site $j$ do {\em not} correspond to an orthogonal basis. Hence,
$(j,\beta'|j,\beta)\neq\delta_{\beta\beta'}$, and we must {\em construct}
a new basis of left states such that $\{j,\beta'|j,\beta)=
\delta_{\beta\beta'}$.
Let $B$ denote the $10\times10$ matrix whose columns are the ten states
$|\beta)\equiv|j,\beta)$, independent of $j$. Then the matrix inverse
$B^{-1}$ is precisely the orthogonal basis that we need, with rows
$\{\beta'|$. In addition to orthogonality one has, by construction,
the closure relations
\eq{3.11}
{\sum_\beta |\beta)\{\beta| = \sum_\beta |\beta\}(\beta| = 1.}
We can bring Eqs. (\ref{3.9}) and (\ref{3.10})
into more symmetric forms by using the identity
\eq{3.12}
{\vop|\alpha)=v_\alpha|\alpha\}.}
Here $v_\alpha$ is the velocity [in the (100) direction] associated with
the state $|\alpha)$.
Eq.~(\ref{3.12}) is a result of (\ref{B.12}) in Appendix B.
Using (\ref{3.12}) in (\ref{3.9}) and (\ref{3.10}) yields
\eqs{3.13}
{T_{\alpha\beta}&\equiv&|t_{\alpha\beta}|^2=
|\frac{4i\hbar}{a_L}\sqrt{v_\beta v_\alpha}\,\{N+1,\beta|G^R|0,\alpha\}|^2, \\
R_{\alpha\beta}&\equiv&|r_{\alpha\beta}|^2=
|\frac{4i\hbar}{a_L}\sqrt{|v_\beta| v_\alpha}\,\{0,\beta|G^R|0,\alpha\}-
\delta_{\alpha\beta}\{0_a,\alpha|0_a,\alpha\}|^2.}

\setcounter{equation}{0}
\section[Current]{$\!\!\!\!\!\!\!.\,\,$ ELECTRIC CURRENT}
\label{sec.current}

\hspace{\parindent}
Transport in solids can be treated on various levels, depending on the
approximations that are assumed at the outset. Examples are the
semiclassical Boltzmann equation, the Kubo formula
for linear response, and the Landauer formula which expresses
the conductance of a system in terms of the single--particle transmission
coefficients \cite{landauer}.

In the case of a perfectly layered heterostructure it is well known that
an expression for the electric current can be written down directly in terms
of the transmission coefficients.
Assume that one can define chemical potentials $\mu_L$ and $\mu_R$ in the left
and right leads, respectively. Under equilibrium conditions $\mu_L=\mu_R$, and
when a bias $V$ is applied between the two leads, one has $\mu_R=\mu_L-eV$.
Strictly speaking, these
potentials do not describe the situation close to the barrier
structure, a region which is out of equilibrium because of tunneling electrons.
Thus, $\mu_L$ and $\mu_R$ correspond to asymptotic distributions in the leads.
In the left lead, states with energy $E$ are occupied with probability
$f_{FD}(E-\mu_L)=[\exp((E-\mu_L)/k_B T)+1]^{-1}$. Here $f_{FD}$ is the
Fermi--Dirac distribution function, $k_B$ is Boltzmann's constant, and $T$ is
the temperature. An electron in initial state $|\alpha)$ with velocity
$v_\alpha$ in the (100) direction has probability $T_{\alpha\beta}$ of being
transmitted into the right lead in a final state $|\beta)$, which has
probability $1-f_{FD}(E-\mu_L+eV)$ of being empty. Analogous arguments hold
for the states in the right lead. In order to find the total current flowing
in the (100) direction, we must integrate over all possible states, i.e., over
the Brillouin zone. In addition we have to sum over the different energy
bands for each $\kvec$.

In the present case we have scattering processes that conserve parallel
momentum and total energy, whereas the momentum $\hbar k$ in the (100)
direction may be altered. It is therefore convenient to replace the integral
over $k$ and the sum over energy bands by an integral over energy and a sum
over momenta $\hbar k_\alpha$:
\eq{4.1}
{\sum_n\int\frac{dk}{2\pi}=\sum_\alpha\int\frac{dE}{2\pi}\left(\frac{\partial
E}{\partial k_\alpha}\right)^{-1}.}
Since $(\partial E/\partial k_\alpha)^{-1}=(\hbar v_\alpha)^{-1}$, the total
electric--current density is given by
\eq{4.2}
{J=-\frac{e}{\hbar}\sum_{\alpha\beta}\int\frac{d\kp}{(2\pi)^2}\int
\frac{dE}{2\pi}T_{\alpha\beta}(E,\kp)
\left[f_{FD}(E-\mu_L)-f_{FD}(E-\mu_L+eV)\right].}
Here $e$ is the absolute value of the electron charge so that $V>0$ results in
a net flow of electrons from left to right, i.e., a negative electric current.
The sum over $\alpha$ and $\beta$ is restricted to states with velocities
$v_\alpha,v_\beta>0$ in the (100) direction, and we have taken advantage of
time--reversal symmetry under which $T(k_\alpha\to k_\beta)=T(-k_\beta\to
-k_\alpha)$. The integral over parallel momentum $\kp$ runs over the {\em
two--dimensional Brillouin zone} which for the [100] plane in zincblende
structures is a square with corners $(0,\pm 2\pi/a_L)$ and $(\pm 2\pi/a_L,0)$.

Note that the simple arguments above are no longer applicable if one wants
to include effects of alloy scattering or electron--phonon interactions.
In that case a full Green--function treatment is required, and in Appendix B
such an approach is presented. For the ideal case, where (\ref{4.2}) is
valid, we demonstrate explicitly the equivalence between the Green--function
approach and the transmission--probability approach.

\setcounter{equation}{0}
\section[Results]{$\!\!\!\!\!\!\!.\,\,$ Results and discussion}
\label{sec.results}

\hspace{\parindent}
The model and method described above can be applied to study electronic
transport through arbitrary heterostructures with zincblende--type
constituents. In this section we shall concentrate on the $\Gamma $X mixing
and transfer, and results will be presented exclusively for GaAs/AlAs
heterostructures. We have used a simple, linear relation,
\eq{5.1}
{\Delta E_c^\Gamma(x)=0.9x \ \ \ (\mbox{eV})\;,}
for the $\Gamma$ point conduction--band offset between Al$_x$Ga$_{1-x}$As
and GaAs. This is a compromise between various empirical relations quoted
in the literature \cite{adachi2}. In combination with the TB
parameters of Ref.~\cite{vogl}
(see Appendix A), Eq. (\ref{5.1}) yields a discontinuity
of about 0.16 eV between $E_c^X$(AlAs), i.e., the $X$ point conduction--band
minimum in AlAs, and $E_c^\Gamma$(GaAs). This is in agreement with recent
experiments \cite{lu1}--\cite{lu2}.

We believe it is instructive to base the following discussion on parts of the
{\em complex band structure} of GaAs and AlAs. In Fig. 2 the solid lines
denote the lowest conduction band in the (100) direction, i.e., for
$\kp=0$. For AlAs we have also plotted the two branches of complex wave
vector corresponding to the analytical continuation below the conduction--band
minima at the $\Gamma$ and $X$ points of the Brillouin zone \cite{xmin}.
These branches are found by solving the Schr\"{o}dinger equation for fixed
energy $E$. In the present model there are 10 solutions $k(E)=k_R+i\kappa$
that make up the full complex band structure. In Fig. 2 we have only plotted
the imaginary parts, $\kappa_\Gamma$ and $\kappa_X$, of the two branches
that are relevant to tunneling in a GaAs/AlAs heterostructure. For an incoming
electron at energy $E$ in the $\Gamma$ valley of GaAs, $\kappa_\Gamma(E)$ and
$\kappa_X(E)$ determine the exponential decay of the tunneling wave functions
$\psi_\Gamma$ and $\psi_X$ through the AlAs barrier(s). Clearly, the relative
importance to the tunneling process of $\psi_\Gamma$ and $\psi_X$ will depend
on which has the slowest decay in the barrier region. For energies below, but
close to the
conduction--band minima we have $\kappa_\Gamma(E)\simeq[2m_\Gamma^*(E_c^\Gamma
-E)/\hbar^2]^{1/2}$ and $\kappa_X(E)\simeq[2m_X^*(E_c^X-E)/\hbar^2]^{1/2}$,
where $m_\Gamma^*$ and $m_X^*$ are the effective masses in the $\Gamma$ and
the $X$ valley, respectively. Thus, although the ``$X$ barrier'' $E_c^X-E$
is much lower then the ``$\Gamma$ barrier'' $E_c^\Gamma-E$, the large value of
$m_X^*$ yields $\kappa_X>\kappa_\Gamma$ for energies up to about 20 meV above
the conduction--band minimum in GaAs. Furthermore it is not
only the exponential decay of the tunneling wave functions that determine
their contribution to transmission through a barrier. At the material
interfaces plane waves in GaAs must be matched to the wave functions
$\psi_\Gamma$ and $\psi_X$ in the AlAs barrier. From symmetry arguments we
expect, for an incoming $\Gamma$ electron,
a larger matrix element for the matching to $\psi_\Gamma$ than to
$\psi_X$. These effects are illustrated in Fig. 3 where we have plotted the
absolute value of the tunneling states as a function of position for a
GaAs/AlAs potential step.
The solid lines represent states at an energy 10 meV above $E_c^\Gamma$(GaAs).
The stronger coupling to $\psi_\Gamma$ is clearly
observed by the fact that $|\psi_\Gamma|\gg|\psi_X|$ in the first layer of
AlAs (i.e., for $N_b=1$).
Furthermore, since $\kappa_X\ga\kappa_\Gamma$ for this energy, $\psi_X$
decays slightly
faster than $\psi_\Gamma$, and tunneling will predominantly happen via
$\psi_\Gamma$. At an energy of 100 meV above $E_c^\Gamma$(GaAs)
(dashed lines) we see again the stronger coupling to $\psi_\Gamma$ at the
material interface. However, in this case $\kappa_X$ is considerably
smaller than $\kappa_\Gamma$, and beyond the 10th AlAs monolayer we have
$|\psi_X|>|\psi_\Gamma|$. Thus, depending on the width of the AlAs barrier,
tunneling at this energy may take place via $\psi_\Gamma$, $\psi_X$ or both.

In a typical experiment the Fermi level is
about 10--50 meV above $E_c^\Gamma$(GaAs). With the aid of Fig. 4 we will
now sketch a qualitative picture of the tunneling of an
electron through a single AlAs barrier,
where a bias $V$ is applied such that transfer to the GaAs $X$ valley is
possible. Let us assume that the energy of the electron is far from
any resonances in the system. At the left interface $(x_1)$ the matching of
the incoming plane wave to tunneling states $\psi_\Gamma$ and $\psi_X$ may
be described with matrix elements $M_{\Gamma\Gamma}(x_1)$ and
$M_{\Gamma X}(x_1)$, respectively. We saw in Fig. 3, in accordance with
expectations, that $M_{\Gamma\Gamma}\gg M_{\Gamma X}$. Since there is a bias
across the barrier, the decay parameters $\kappa_\Gamma$ and $\kappa_X$ will
now be functions of position within the barrier. Between $x_1$ and $x_2$
(typically a very short distance), we have $\kappa_X>\kappa_\Gamma$, so up to
this point tunneling via $\psi_\Gamma$ dominates strongly. However, beyond
$x_2$ we have $\kappa_X<\kappa_\Gamma$, and the amplitude for tunneling
via $\psi_X$ is ``catching up'' relative to that of $\psi_\Gamma$. Between
$x_3$
and $x_4$ the energy of the tunneling electron lies above $E_c^X$(AlAs). This
means that $\kappa_X=0$, and the amplitude of $\psi_X$ is not further reduced
whereas $\psi_\Gamma$ continues to decay exponentially. Finally, at the
right interface $(x_4)$ matching of the two tunneling states to transmitted
states in the collector contact may be described by matrix elements
$M_{\alpha\beta}(x_4)$, with $\alpha,\beta=\Gamma$ or $X$. As above we expect
the coupling between states of different symmetry $(M_{\Gamma X}$ and
$M_{X\Gamma})$ to be much weaker than that between ``similar'' states
$(M_{\Gamma\Gamma}$ and $M_{XX})$.

Based on this simple picture we can make predictions concerning the two
contributions, $J_\Gamma$ and $J_X$, to the total current density in a
GaAs/AlAs heterostructure. First, there is no $\Gamma X$ transfer for
biases $V<[E_c^X($GaAs$)-E_F]/e$. However, the direct component $J_\Gamma$
may still be strongly affected by the presence of the tunneling  state
$\psi_X$, in particular due to resonant tunneling via confined $X$ states
in the barrier. We will briefly come back to this below. For biases
$V>[E_c^X($GaAs$)-E_F]/e$, $\Gamma X$ transfer is possible, and the total
current density is given by the sum of $J_\Gamma$ and $J_X$. Their relative
contribution is determined by two competing factors. On one hand,
$J_\Gamma$ is favored by the stronger coupling to $\psi_\Gamma$ at the
barrier interface. On the other hand, $J_X$ is favored by the slower
exponential decay of $\psi_X$ through the barrier. For very thin barriers
the interface coupling is the decisive factor, and we expect
$J_\Gamma>J_X$. With increasing thickness of the barrier the difference
in decay rate will eventually yield $J_X>J_\Gamma$, and the crossover to a
``$\Gamma X$ filter'' takes place around a barrier width
$L_{\Gamma X}$ that will be estimated below via numerical examples.

The qualitative discussion above can only be expected
to be valid in the off--resonant
tunneling regime. The resonant features of tunneling through indirect--bandgap
heterostructures have been discussed extensively in the literature
\cite{schulz,ting}, and they have been observed very clearly in experiments
\cite{pritchard,austing}. In general there are
two types of resonant states in a
GaAs/AlAs structure. With two or more AlAs barriers there will be one or more
GaAs wells in which one finds the ``normal'' quasibound states
$\eps_\Gamma$ confined by the $\Gamma$ profile of the conduction
band. On the other hand, the $X$ profile of the conduction band reverses
the role of barrier and well material, and as a result there will be
resonant levels $\eps_X$ confined to the AlAs layers. These states
are degenerate with the continuum of the GaAs contacts and hence similar
to the socalled Fano resonances in atomic physics \cite{fano}.
Predictions concerning
resonance positions and linewidths are to some extent model dependent, as
discussed in Ref.~\cite{ting}. Here we shall not elaborate further
on the resonant structure of the current,
although tunneling via resonant levels is indeed taking place in the structures
studied below ({\em only} $\eps_X$ in the single--barrier case, {\em both}
$\eps_\Gamma$ and $\eps_X$ in the double--barrier case).

In order to determine the conditions for having a ``$\Gamma X$ filter'', we
have calculated current--voltage curves for a number of single-- and
double--barrier structures. Explicit results will be shown for single
barriers only.
We have used a Fermi level of 10 meV above
the GaAs conduction--band edge. Furthermore the potential profile is
taken to be flat throughout the GaAs contacts, with a linear voltage drop
across the barrier structure. In Fig. 5 we have
plotted the two current--density
components $J_\Gamma$ (thick solid line) and $J_X$ (thick dashed line)
for a single AlAs barrier of width 40 \AA, for biases in the range where
$\Gamma X$ transfer is possible \cite{belowx3}, and at zero temperature.
We have also included
the corresponding transmission coefficients $T_\Gamma$ (thin solid line)
and $T_X$ (thin dashed line) at the Fermi level and with $\kp=0$. This shows
that, although $J_\Gamma$ and $J_X$ are given by integrals over $E$ and $\kp$
of $T_\Gamma$ and $T_X$, respectively, their dependence on applied bias is to
a very good approximation reflected in a {\em single value} of the
transmission coefficient. Of course, the integrals over $E$ and $\kp$ tend
to give a smoother behaviour of $J(V)$ than that of $T(V)$. In addition,
resonance positions and relative peak values may be slightly shifted in
$J(V)$ when compared with $T(V)$. However, since we are not interested in
details in the current--voltage curves, the following discussion may be based
on the behaviour of $T_\Gamma$ and $T_X$ at $E=E_F$ and $\kp=0$.

{}From Fig. 5 we see that a single AlAs barrier acts like a good $\Gamma X$
filter for a width of 40 \AA. As a measure of the $\Gamma X$ transfer we have
calculated the ratio
\eq{5.2}
{R_{\Gamma X}\equiv\frac{\int dV\,T_X(V;E=E_F;\kp=0)}
{\int dV\,T_\Gamma(V;E=E_F;\kp=0)}\;,}
with limits of integration at 0.5 and 0.8 V \cite{rinterval}. The result
is shown in Fig. 6 where $R_{\Gamma X}$ is plotted versus the number of
monolayers of AlAs in the barrier. As expected there is only negligible
$\Gamma X$ transfer for very narrow barriers, but already at 10 monolayers
there is almost an order of magnitude more electrons being
transmitted in the $X$ valley than in the $\Gamma$ valley. Hence, the single
AlAs barrier behaves like a $\Gamma X$ filter with a crossover thickness
$L_{\Gamma X}\sim 25$\AA.

We have also performed analogous calculations for double--barrier structures.
As mentioned above,
a qualitative difference from the single--barrier case is that resonant
tunneling may happen not only via confined $X$ states in the barriers, but also
via confined $\Gamma$ states in the well. Hence the resulting current--voltage
curves display more resonances, both asymmetric Fano resonances and ``normal''
resonances with a Breit--Wigner form.

Because of the complicated resonant behaviour of $T_{\Gamma}$ and $T_X$,
$R_{\Gamma X}$ is not a smoothly increasing function of barrier width. This
is already apparent in Fig. 6, and in the case of a double barrier the
oscillations are even stronger. Even so it is possible to distinguish two
regimes, one where mostly $R_{\Gamma X}\ll 1$ and another where mostly
$R_{\Gamma X}\gg 1$. For the double--barrier structure
we have evaluated $R_{\Gamma X}$ as a function of AlAs barrier width,
for four different GaAs well widths: 6, 17, 34 and 56 \AA. The
resulting barrier crossover widths are $L_{\Gamma X}\sim14,22,25$ and
25 \AA, respectively. This result can be understood as follows. When the well
is very narrow and consists of only two monolayers of GaAs, there is strong
coupling between the $X$ states in each barrier. Hence the crossover width
for one barrier is about half the crossover width in the single--barrier case.
With increasing well width the coupling between the $X$ states in each
barrier becomes weaker, and $L_{\Gamma X}$ approaches the value found in the
single--barrier case.

\setcounter{equation}{0}
\section[Conclusion]{$\!\!\!\!\!\!\!.\,\,$ Conclusions}
\label{sec.concl}

\hspace{\parindent}
We have presented a framework for the study of tunneling in III--V
semiconductor heterostructures. A multiband tight--binding model was used
to obtain a realistic description of the lowest conduction band of each
material. We applied the model to GaAs/AlAs structures where the barrier
material has an indirect bandgap. Tunneling states $\psi_{\Gamma}$ and
$\psi_X$ contribute to transmission. For barriers wider than about ten
monolayers of AlAs, tunneling via $\psi_X$ dominates. An applied bias larger
than, in the present model, 0.5 V enables transfer of $\Gamma$ electrons
to the $X$ valley in GaAs. The $\Gamma X$ transfer is stimulated by an
indirect--bandgap AlAs barrier. With a conduction--band offset of 160 meV
between GaAs and AlAs we find a barrier ``crossover width''
$L_{\Gamma X}\sim25$ \AA\ which separates a regime of negligible
$\Gamma X$ transfer (narrow barriers) from a regime of large
$\Gamma X$ transfer (wide barriers).

Finally we would like to comment that the present model is quite
general. Given the crystal structure, a set of tight--binding parameters,
and the conduction--band offsets between the materials involved, one can
study transport through structures of arbitrary composition. Interesting
applications, besides the one studied here, could be the polytype type--II
heterostructures \cite{soderstrom} and the mixing of hole states in a
$pn$ junction. One might also try to apply the model on the level of the
coherent--potential approximation, first, to enable a study of the effects
of alloy scattering in e.g. Al$_x$Ga$_{1-x}$As, and second, to see how
interface roughness affects the $\Gamma X$ transfer.

\vspace{0.5cm}
\renewcommand{\thesection}{ACKNOWLEDGMENT}
\section[Acknowl]{}
\hspace{\parindent}
This work was supported in part by Norges Forskningsr\aa d.

\vspace{0.5cm}
\renewcommand{\thesection}{APPENDIX A}
\setcounter{section}{0}
\renewcommand{\theequation}{\Alph{section}\arabic{equation}}
\setcounter{equation}{0}
\section[App.A]{}


\hspace{\parindent}
{}From Eq.~(\ref{2.1}) we have
\eq{A1}
{(nbj\kp|H|n'b'j'\kp)=L_{BZ}^{-1}\int dk_x e^{ik_x(j-j')a_L/4}
 (nb\kvec|H|n'b'\kvec),}
where the matrix elements $(nb\kvec|H|n'b'\kvec)$ are given in table A of
Ref.~\cite{vogl}. For example, the matrix element between the anion $s$
orbital and the cation $p_z$ orbital is
\eq{A2}
{(sa\kvec|H|p_zc\kvec)=V(sa,pc)g_3(\kvec).}
Here $V(sa,pc)=4(sa\Rvec_a|H|pc\Rvec_c)$ is one of the, thirteen in all,
independent TB matrix elements in the basis of symmetrically
orthogonalized atomic orbitals $|nb\Rvec)$, and $g_3(\kvec)$ may be
regarded as a form factor that is determined by the symmetry of the
crystal \cite{harrison}. One may write
\eq{A3}
{g_3(\kvec)=e^{ik_xa_L/4}\frac{i}{2}\sin\kp\dvec_1-
 e^{-ik_xa_L/4}\frac{i}{2}\sin\kp\dvec_2,}
where $\dvec_1=(1,1)a_L/4$ and $\dvec_2=(1,-1)a_L/4$
are vectors in the $(y,z)$ plane. Hence (\ref{A1}) readily yields
\eq{A4}
{(saj\kp|H|p_zcj'\kp)=V(sa,pc)\left[\delta_{j+1,j'}\frac{i}{2}\sin\kp\dvec_1-
 \delta_{j-1,j'}\frac{i}{2}\sin\kp\dvec_2\right]}
for the matrix element in the desired basis. In Eq.~(\ref{2.2}) we let
$U_{ac}$ represent the ``hopping'' from an anion layer $j$ to a cation
layer $j-1$, whereas $V_{ac}$ represents the hopping in the other direction,
from $j$ to $j+1$ \cite{labeling}.
We then have, with $s_n\equiv\frac{1}{2}\sin\kp\dvec_n$
and $c_n\equiv\frac{1}{2}\cos\kp\dvec_n$
$(n=1,2)$, the following elements of our Hamiltonian:
\eqs{A5}
{
E_b&=&
\left[ \begin{array}{ccccc}
                     E(s,b) & & & &        \\
                     & E(p,b) & & &        \\
                     & & E(p,b) & &        \\
                     & & & E(p,b) &        \\
                     & & & & E(s^*,b)
\end{array} \right] \ \ ; \ \ b=c\; \mbox{or}\;a\;,     \\
U_{ac}&=&
\left[ \begin{array}{ccccc}
 V(s,s)c_2 & -V(sa,pc)c_2 & iV(sa,pc)s_2 & -iV(sa,pc)s_2 & 0              \\
 V(pa,sc)c_2 & V(x,x)c_2 & -iV(x,y)s_2 & iV(x,y)s_2 & V(pa,s^*c)c_2       \\
 -iV(pa,sc)s_2 & -iV(s,y)s_2 & V(x,x)c_2 & -V(x,y)c_2 & -iV(pa,s^*c)s_2   \\
 iV(pa,sc)s_2 & iV(x,y)s_2 & -V(x,y)c_2 & V(x,x)c_2 & iV(pa,s^*c)s_2      \\
 0 & -V(s^*a,pc)c_2 & iV(s^*a,pc)s_2 & -iV(s^*a,pc)s_2 & V(s^*,s^*)c_2
\end{array} \right],   \\
V_{ac}&=&
\left[ \begin{array}{ccccc}
 V(s,s)c_1 & V(sa,pc)c_1 & iV(sa,pc)s_1 & iV(sa,pc)s_1 & 0              \\
 -V(pa,sc)c_1 & V(x,x)c_1 & iV(x,y)s_1 & iV(x,y)s_1 & -V(pa,s^*c)c_1       \\
 -iV(pa,sc)s_1 & iV(x,y)s_1 & V(x,x)c_1 & V(x,y)c_1 & -iV(pa,s^*c)s_1   \\
 -iV(pa,sc)s_1 & iV(x,y)s_1 & V(x,y)c_1 & V(x,x)c_1 & -iV(pa,s^*c)s_1      \\
 0 & V(s^*a,pc)c_1 & iV(s^*a,pc)s_1 & iV(s^*a,pc)s_1 & V(s^*,s^*)c_1
\end{array} \right].  }
In addition $U_{ca}=U_{ac}^{\dag}$ and $V_{ca}=V_{ac}^{\dag}$.
Numerical values for GaAs and AlAs, taken from Ref.~\cite{vogl}, are
collected in Table 1.
\begin{table}
\begin{tabular}{||l|rrrrrrr||}
\hline\hline
Compound & $E(s,a)$ & $E(p,a)$ & $E(s,c)$ & $E(p,c)$ & $E(s^*,a)$ &
$E(s^*,c)$ & $V(s,s)$ \\
\hline
GaAs & -8.3431 & 1.0414 & -2.6569 & 3.6686 & 8.5914 & 6.7386 & -6.4513 \\
AlAs & -7.5273 & 0.9833 & -1.1627 & 3.5867 & 7.4833 & 6.7267 & -6.6642 \\
\hline\hline
Compound & $V(x,x)$ & $V(s,y)$ & $V(sa,pc)$ & $V(pa,sc)$ & $V(s^*a,pc)$ &
$V(pa,s^*c)$ & $V(s^*,s^*)$ \\
\hline
GaAs & 1.9546 & 5.0779 & 4.4800 & 5.7839 & 4.8422 & 4.8077 & 0.0000 \\
AlAs & 1.8780 & 4.2919 & 5.1106 & 5.4965 & 4.5216 & 4.9950 & 0.0000 \\
\hline\hline
\end{tabular}
\caption{Empirical matrix elements of the $sp^3s^*$ Hamiltonian in eV
for GaAs and AlAs.}
\end{table}

\vspace{0.5cm}
\renewcommand{\thesection}{APPENDIX B}
\renewcommand{\theequation}{\Alph{section}\arabic{equation}}
\setcounter{equation}{0}
\section[App.B]{}


\hspace{\parindent}
In this Appendix we will follow a general approach and obtain the
electric--current density in terms of nonequilibrium Green functions.
Using socalled ``surface Green functions'' we will show how to make
contact with the transmission--probability approach presented in
Sec. IV.  The present treatment is closely related to the one
introduced by Caroli {\em et al.} \cite{caroli}.
For a system of noninteracting electrons, the total current density
$J$ is given in terms of the single--particle current--density
operator $\hat{J}$
as a many--body {\em quantum} average over occupied single--particle
states, and a {\em statistical} average over all possible many--body
configurations. Thus,
\eq{B.1}
{J=\sum_{\{m\}}P(m)\sum_s p_m(s)(\phi_s|\hat{J}|\phi_s),}
where $P(m)$ is the probability of having the many--body configuration
$\{m\}$, and $p_m(s)$ is the probability that the single--particle
state $|\phi_s)$ is occupied in the given many--body configuration
[$p_m(s)$ equals 0 or 1]. Since the {\em reduced density matrix} $\rho$
is defined as
\eq{B.2}
{\rho=\sum_{\{m\}}P(m)\sum_s p_m(s)|\phi_s)(\phi_s|,}
we may also write
\eq{B.3}
{J=\tr(\rho\hat{J}),}
where $\tr$ denotes the trace operation. The connection to Green functions
is now transparent since
\eq{B.4}
{\rho=\int\frac{dE}{2\pi}\GL(E),}
where we follow the notation in e.g. Ref.~\cite{lipavsky}.
The {\em correlation}
function $\GL$ and its hole ``counterpart'' $\GG$ are related to the
{\em retarded} and {\em advanced} Green functions $G^R$ and $G^A$ via
\cite{lipavsky}
\eq{B.5}
{\GL+\GG=i(G^R-G^A).}
In general only two of the four functions in (\ref{B.5}) are independent
since one also has the connecting identity
\eq{B.6}
{G^R=\left[G^A\right]^{\dag}.}
Eqs. (\ref{B.3}) -- (\ref{B.6}) constitute the proper starting point for
evaluation of the electric--current density
in a system of noninteracting electrons.

An operator $\Jop$ for the electric--current density is found by comparing
the continuity equation and the Heisenberg equation of motion. We also
use (\ref{B.4}) and obtain the result
\eq{B.7}
{J_{m,m+1}=-\frac{e}{\hbar}\int\frac{dE}{2\pi}\int\frac{d\kp}{(2\pi)^2} i
\tr (\GL_{m+1,m}u_{m,m+1}-\GL_{m,m+1}u_{m+1,m})}
for the net current density flowing between layers $m$ and $m+1$ (in the
positive $x$ direction). In (\ref{B.7}) the matrix elements of $\GL$
depend, of course, on $E$ and $\kp$, and the TB ``hopping'' elements
$u_{m,m\pm1}$ depend on $\kp$ in a way which is determined by the crystal
structure (see Appendix A). Eq.~(\ref{B.7}) is valid for systems that have
translational invariance in the parallel plane and are characterized by
a nearest--neighbour TB Hamiltonian.

In order to connect the Green--function approach and the
transmission--probability approach, we must show that
\eq{B.8}
{i\tr(\GL_{m+1,m}u_{m,m+1}-\GL_{m,m+1}u_{m+1,m})=\sum_{\alpha\beta}
T_{\alpha\beta}\left[f_{FD}(E-\mu_L)-f_{FD}(E-\mu_L+eV)\right]}
when we have a structure connected to semiinfinite perfect leads in equilibrium
at chemical potentials $\mu_L$ and $\mu_R=\mu_L-eV$. The following ingredients
are required to rewrite the left--hand side of this equation:
\begin{itemize}
\item
The equilibrium expressions for $\GL$ and $\GG$ in a homogeneous system
at chemical potential $\mu$ \cite{kadanoff}:
\eqs{B.9}
{g^<(E)&=&i\left[g^R(E)-g^A(E)\right]f_{FD}(E-\mu), \\
 g^>(E)&=&i\left[g^R(E)-g^A(E)\right]\left[1-f_{FD}(E-\mu)\right].}
(We use lowercase letters for Green functions that describe a system
in equilibrium.)
\item
The rules, derived by Langreth and Wilkins \cite{langreth}, for handling
products of two or more Green functions:
\eqs{B.10}
{(AB)^R    &=& A^R B^R,  \\
 (AB)^A    &=& A^A B^A, \\
 (AB)^{<}  &=& A^R B^{<} + A^{<} B^A, \\
 (AB)^{>}  &=& A^R B^{>} + A^{>} B^A.}
\item
The Dyson equation and the recursive Green--function technique, described
in detail in the Appendix of Ref.~\cite{stovneng}.
\item
The unit matrices of Eq.~(\ref{3.11}).
\end{itemize}
Straightforward algebra then yields
\eqar{B.11}
{& &\sum_{\alpha\beta\mu\nu}(\alpha_c|U_{ca}2\mbox{Im}\gamma^+_a
U_{ac}|\beta_c)
\{\beta|G^R_{N+1,0}|\mu\}(\mu_a|U_{ac}2\mbox{Im}\gamma^-_c U_{ca}|\nu_a)
\times \nonumber \\
& &\;\;\;\;\;\;\{\nu|G^A_{0,N+1}|\alpha\}[f_{FD}(E-\mu_L)-f_{FD}(E-\mu_L+eV)] }
for the left--hand side of (\ref{B.8}), where we have chosen to let the
cation and anion layers of ``site'' $N+1$ represent $m$ and $m+1$,
respectively. Here subscripts $c$ or $a$ on the states $\alpha,\beta,\mu$ and
$\nu$ denote the cation or anion parts, respectively. Further, 2Im$\gamma\equiv
i(\gamma^A-\gamma^R)$ \cite{imagmatrix}, and $\gamma^+_a (\gamma^-_c)$ is the
surface Green function, i.e., the local element at the surface layer of the
Green function for  a semiinfinite crystal extending from layer
$N+1,a$ ($0,c$) to $+\infty$ ($-\infty$). Remember, sites $N+1$ and $0$ are
defined as the first sites of the semiinfinite crystals at {\em constant}
potential energy on each side of the heterostructure. From the recurrent
relations for the surface Green functions \cite{stovneng} it is clear
that $\gamma^+_a$ is the same on site $N+1$ as in $+\infty$, and $\gamma^-_c$
the same on site $0$ as in $-\infty$. In other words, $\gamma^+_a$ and
$\gamma^-_c$ in (\ref{B.11}) reflect asymptotic properties of the left and
right lead, respectively, which allows for the use of (\ref{B.9}) in deriving
(\ref{B.11}).

In (\ref{B.11}) only a few terms contribute to the sum.
First, there can only be a contribution from
{\em extended} states with real values of $k$. Clearly, the result of
(\ref{B.11}) must be unchanged if we choose sites, other than $N+1$ and 0,
further into the asymptotic regions. However, {\em evanescent} states
$\beta$ or $\mu$ would give a modulus $|\{\beta|G^R_{ij}|\mu\}|$ that
decreases exponentially with increasing $i$ or decreasing $j$
($i\geq N+1$, $j\leq 0$). Second, only states with {\em positive}
velocity contribute to the sum since the {\em retarded} Green function
propagates a scattering state forward in time.
Finally, since $G^A=[G^R]^{\dag}$, the states $\nu$ and $\alpha$ must
also be extended states with positive velocities.

What remains is to evaluate the matrix elements
$(\alpha_c|U_{ca}2\mbox{Im}\gamma^+_a U_{ac}|\beta_c)$ and
\newline
$(\mu_a|U_{ac}2\mbox{Im}\gamma^-_c U_{ca}|\nu_a)$. This task is accomplished
by rewriting matrix elements of the velocity operator $\vop$
[see Eq.~(\ref{3.5})] with help of the Dyson equation and the recursive
Green--function technique. The result is
\equthree{B.12}
{(\alpha_c|U_{ca}2\mbox{Im}\gamma^+_a U_{ac}|\beta_c)&=&-\frac{4\hbar}{a_L}
(\alpha|\vop|\beta)&=&-\frac{4\hbar}{a_L}v_{\alpha}\delta_{\alpha\beta}
 \ \ \ (v_{\alpha}>0), \\
(\mu_a|U_{ac}2\mbox{Im}\gamma^-_c U_{ca}|\nu_a)&=&\frac{4\hbar}{a_L}
(\mu|\vop|\nu)&=&\frac{4\hbar}{a_L}v_{\mu}\delta_{\mu\nu} \;\;\;\;(v_{\mu}<0).}
Here the relation $(\alpha|\vop|\beta)=v_{\alpha}\delta_{\alpha\beta}$
follows from current conservation. Although states with negative velocity
do not contribute to the sum in (\ref{B.11}), there corresponds
to each state $\mu$ (with $v_{\mu}<0$) a ``time--reversed'' state
$\mu'$ (with $v_{\mu'}=-v_{\mu}>0$) for which one has \cite{vrelation}
\eq{B.13}
{(\mu'_a|U_{ac}2\mbox{Im}\gamma^-_c U_{ca}|\mu'_a)=
-\frac{4\hbar}{a_L}v_{\mu'}\;\;.}
Finally, collecting all our knowledge, the sum in (\ref{B.11}) may be
written as
\eqar{B.14}
{&\;&\sum_{\alpha\beta\mu\nu}\frac{4\hbar}{a_L}v_{\alpha}\delta_{\alpha\beta}
\{\beta|G^R_{N+1,0}|\alpha\}\frac{4\hbar}{a_L}v_{\mu}\delta_{\mu\nu}
\{\nu|G^A_{0,N+1}|\alpha\}[f_{FD}(E-\mu_L)-f_{FD}(E-\mu_L+eV)] \nonumber \\
&=&\sum_{\alpha\beta} \left( \frac{4\hbar}{a_L} \right)^2
v_{\alpha}v_{\beta}|\{\beta|G^R_{N+1,0}|\alpha\}|^2
[f_{FD}(E-\mu_L)-f_{FD}(E-\mu_L+eV)] \nonumber \\
&\equiv& \sum_{\alpha\beta} T_{\alpha\beta}
[f_{FD}(E-\mu_L)-f_{FD}(E-\mu_L+eV)]\;,}
which demonstrates the validity of Eq.~(\ref{B.8}),
and hence the equivalence of the
transmission--probability approach in Sec.~\ref{sec.current} and the
Green--function approach.

\vspace{2cm}
\noindent
$^*$ Present address: Institutt for fysikk, Norges Tekniske
H\o gskole, Universitetet i Trondheim, N--7034 Trondheim--NTH, Norway.

\vspace{0.5cm}
\begin{Large}
\begin{bf}
\noindent
Figure Captions
\end{bf}
\end{Large}

\vspace{.5cm}
\noindent
FIG.1. Projection of the zincblende lattice on the (100) direction. Solid
and open circles represent cation and anion layers, respectively. The
interlayer distance is $a_L/4$, i.e., one forth the lattice constant.
Indicated are also the ``on--layer'' energies $E_c$ and $E_a$, and the
interlayer hopping matrices $U_{ca}, U_{ac}, V_{ca}$ and $V_{ac}$.

\vspace{.5cm}
\noindent
FIG.2. Parts of the complex band structure for GaAs and AlAs. The solid
lines are the lowest conduction band of the two materials. For AlAs we have
also included the imaginary parts, $\kappa_{\Gamma}$ and $\kappa_X$, of the
analytical continuation below the minima at the $\Gamma$ and $X$ points,
respectively. [To be precise, the conduction--band minimum of AlAs is not
exactly at the $X$ point in the present model, but slightly below, at
$k\simeq 0.85\times(2\pi/a_L)$.] Zero energy is taken to be at the top
of the valence band in GaAs, and with the present model and parameters GaAs
has an energy gap of 1.55 eV.

\vspace{.5cm}
\noindent
FIG.3. Spatial decay into AlAs of the wave functions $\psi_{\Gamma}$
(thick lines) and $\psi_X$ (thin lines) at energies 10 meV (solid lines)
and 100 meV (dashed lines) above
the GaAs conduction--band edge $E_c^{\Gamma}$(GaAs). The distance from the
GaAs/AlAs interface is given in terms of the number of AlAs monolayers $N_b$.
The decay lengths corresponding to these curves are $\kappa_{\Gamma}^{-1}=
0.95a_L$ and $0.98a_L$ ; $\kappa_X^{-1}=0.92a_L$ and $1.78a_L$.
Here the first (second) number refers to the solid (dashed) lines,
and $a_L=5.66$ \AA\ is the lattice constant in AlAs. The inset illustrates
the $\Gamma$ point conduction--band minimum which represents a potential
step between $N_b=0$ and $N_b=1$. The wave functions are normalized to an
incoming plane wave of unit amplitude (see Sec. III).

\vspace{.5cm}
\noindent
FIG.4. Tunneling of an electron through a single AlAs barrier with applied
bias $V$. The profiles of the $\Gamma$ and $X$ point conduction--band
minima are represented by the solid and dashed lines, respectively.
The lowest conduction band in GaAs [in the (100) direction] is sketched
on both sides of the barrier, and the dotted portions indicate initially
occupied states on the left side and available transmission states on the
right side. Shaded areas denote the filled equilibrium Fermi sea in the
emitter and collector contact.
The transmitted electron may contribute to the ``direct'' current density
$J_{\Gamma}$ or the ``transfered'' one $J_X$. The tunneling process
is described in detail in the text. Relaxation processes that bring the
transmitted electrons to thermal equilibrium are indicated with
scattering rates $\tau_{\Gamma\Gamma}^{-1}$ and $\tau_{X\Gamma}^{-1}$.

\vspace{.5cm}
\noindent
FIG.5. Direct and transfered current--density contributions (thick lines),
$J_{\Gamma}$ and $J_X$, for a single AlAs barrier of width 40 \AA. The
corresponding transmission coefficients (thin lines), $T_{\Gamma}$ and $T_X$,
at
$E=E_F=10$ meV and $\kp=0$ reflect to a good approximation both the qualitative
behaviour of $J_{\Gamma}$ and $J_X$, and also their relative contributions
to the total current density. The left vertical axis represents
$J_{\Gamma}$ and $J_X$ in A/cm$^2$; the right axis represents
$T_{\Gamma}$ and $T_X$ which are of the order of $10^{-4}$ in this case.

\vspace{.5cm}
\noindent
FIG.6. $R_{\Gamma X}$, as defined in Eq. (5.2),
for a single AlAs barrier as function of barrier width,
the latter given in terms of $N_b$, the number of monolayers of AlAs.
The solid line connecting the data points is nothing more than a guide
to the eye.
$R_{\Gamma X}$ gives a quantitative measure of the relative contribution
of $J_X$ and $J_{\Gamma}$ to the total current density: $R_{\Gamma X}\ll 1$
implies negligible $\Gamma X$ transfer; $R_{\Gamma X}\gg 1$ implies large
$\Gamma X$ transfer.


\begin{thebibliography}{99}
\bibitem{xmin} In the present model the conduction--band minimum in AlAs
is not exactly at the $X$ point, see Fig. 2.
\bibitem{belowx3} $\Gamma X$ transfer is also possible for biases above
0.8 V. However, the accuracy of the band structure in the
$sp^3s^*$ model diminishes at higher
energies, and it seems reasonable to restrict the presentation to the
lowest conduction band. In this model the second conduction band in GaAs
starts at $X_3^c$, 0.83 eV above $E_c^{\Gamma}$.
\bibitem{rinterval} These integration limits are chosen such that
$R_{\Gamma X}$ reflects a sizable bias interval in which $\Gamma X$
transfer is possible. At the same time we exclude from $R_{\Gamma X}$ the
often large variations of $T_{\Gamma}$ and $T_X$ in the vicinity of the
band edges $X_1^c$ and $X_3^c$ in the GaAs collector contact (i.e.,
$V\ge0.47$ V and $V\le0.83$ V, respectively).
\bibitem{labeling} Here, as well as in Sec. II, $j$ labels individual
atomic layers, in contrast to Sec. III where a common label is used for
pairs of cation and anion layers.
\end{thebibliography}
\end{document}